\documentclass[aps,prb,amsmath,showpacs,twocolumn]{revtex4}
\usepackage{bm}
\usepackage{graphicx}

 	\newcommand{\textmarkright}{%
	 to be published in Appl.\ Phys.\ Lett.}
\begin{document} 
\title{Hysteretic ac loss of polygonally arranged superconducting strips \\
	carrying ac transport current}

\author{Yasunori Mawatari}
\affiliation{%
	National Institute of Advanced Industrial Science and Technology (AIST), 
	Tsukuba, Ibaraki 305--8568, Japan
}
\author{Kazuhiro Kajikawa}
\affiliation{%
	Research Institute of Superconductor Science and Systems, 
	Kyushu University, 
	Fukuoka 819--0395, Japan
}
\date{December 21, 2007}

\begin{abstract}
The hysteretic ac loss of a current-carrying conductor in which multiple superconducting strips are polygonally arranged around a cylindrical former is theoretically investigated as a model of superconducting cables. 
Using the critical state model, we analytically derive the ac loss $Q_n$ of a total of $n$ strips. 
The normalized loss $Q_n/Q_1$ is determined by the number of strips $n$ and the ratio of the strip width $2w$ to the diameter $2R$ of the cylindrical former. 
When $n\gg 1$ and $w/R\ll 1$, the behavior of $Q_n$ is similar to that of an infinite array of coplanar strips. 
\end{abstract}
\maketitle
  	\thispagestyle{myheadings}\markright{\textmarkright}

High-temperature superconducting wires have strip geometry, and in superconducting power transmission cables, the wires are polygonally arranged around a cylindrical former. 
Hysteretic ac loss is a critical parameter of superconducting wires used in these power cables. 
The hysteretic ac loss of a single isolated superconducting strip carrying an ac transport current $Q_1$ was theoretically investigated by Norris~\cite{Norris70} on the basis of the critical state model.~\cite{Bean62} 
Because superconducting cables contain multiple superconducting wires, the electromagnetic interaction among those wires must be taken into account. 
Numerical calculation of ac losses of superconducting cables has been extensively reported,~\cite{Inada04,Sato05,Inada05a,Inada05b,Klincok06,Fukui06,Sato06,Inada06,Kajikawa06} but no analytical investigation has been reported to date. 

In the present letter, we consider polygonally arranged superconducting strips that carry ac transport current as a model of power transmission cables with single-layer structure. 
On the basis of the critical state model, we then derive analytical expressions of hysteretic ac losses in polygonally arranged strips. 

\begin{figure}[b]
\includegraphics{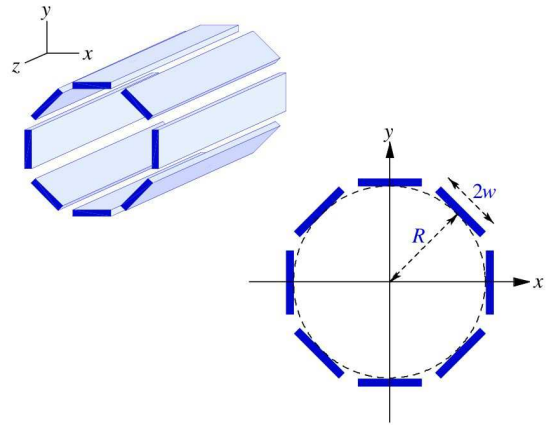}
\caption{(Color online) %
Configuration of polygonally arranged strips, in which $n$ strips of width $2w$ are periodically arranged around a cylindrical former of radius $R$. 
Parameters, $n$, $w$, and $R$, must satisfy $w/R<\tan(\pi/n)$ to avoid intersection of the strips. 
Figure shows $n=8$ as an example. 
}
\label{Fig_polygon}
\end{figure}

We consider superconducting strips that have infinite length along the $z$ axis, and have flat rectangular cross sections ($2w\times d$) in the $xy$ plane. 
These strips are polygonally arranged around a cylindrical former of radius $R$, as shown in Fig.~\ref{Fig_polygon}. 
The thickness $d$ of each strip is much smaller than the width $2w$, and the critical current density $j_c$ is assumed to be uniform and constant as in the Bean model.~\cite{Bean62} 
In such two-dimensional geometry, the current density has only a $z$ component and the magnetic field has $x$ and $y$ components. 
We introduce the complex magnetic field,~\cite{Beth66,Clem73,Zeldov94,Mawatari01,Brojeny02,Sonin02,Mawatari03,Mawatari06a,Clem06} ${\cal H}(\zeta)= H_y(x,y)+iH_x(x,y)$, as a function of the complex variable, $\zeta =x+iy$. 

\begin{figure}[tb]
\includegraphics{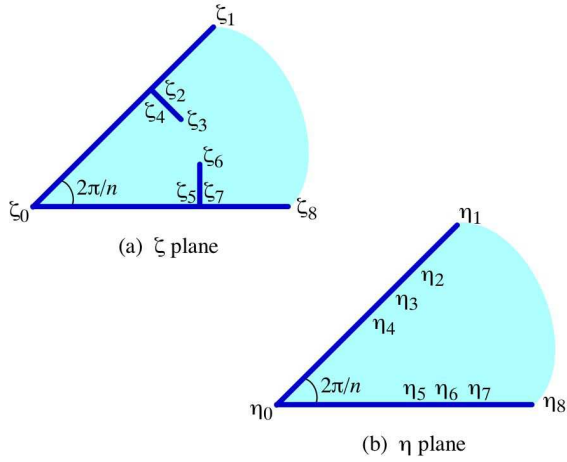}
\caption{(Color online) %
Two complex planes for the conformal mapping given by Eq.~\eqref{zeta-eta}: 
(a) $\zeta=x+iy$ plane and (b) $\eta$ plane. 
The region $0<\arg(\zeta)<2\pi/n$ with two cuts of length $w$ in the $\zeta$ plane is mapped onto the wedge-shaped region of $0<\arg(\eta)<2\pi/n$ in the $\eta$ plane. 
Points $\zeta=\zeta_k$ [where $\zeta_0=0$, $\zeta_2=(R+\epsilon)e^{2\pi i/n}$, $\zeta_3=(R-iw)e^{2\pi i/n}$, $\zeta_4=(R-\epsilon)e^{2\pi i/n}$, $\zeta_5=R-\epsilon$, $\zeta_6=R+iw$, $\zeta_7=R+\epsilon$, and $\epsilon\to +0$ is the positive infinitesimal] are respectively mapped onto points $\eta=\eta_k$ [where $\eta_0=0$, $\eta_2=\alpha e^{2\pi i/n}$, $\eta_3=\gamma e^{2\pi i/n}$, $\eta_4=\beta e^{2\pi i/n}$, $\eta_5=\beta$, $\eta_6=\gamma$, and $\eta_7=\alpha$]. 
}
\label{Fig_cfm-mapping}
\end{figure}

First, we consider the complex field for polygonally arranged strips carrying a transport current $I_0$. 
The complex field ${\cal H}(\zeta)$ in the ideal Meissner state is obtained by the conformal mapping as follows:
\begin{equation}
	{\cal H}(\zeta) = \frac{I_0}{2\pi} \frac{n\eta^{n-1}}{\eta^n -\gamma^n} , 
\label{H(z)_Meissner}
\end{equation}
where the relationship between the complex variables, $\zeta$ and $\eta$, is given by 
\begin{equation}
	\zeta= \int_0^{\eta} d\eta_0 \frac{\eta_0^n-\gamma^n}{%
		\sqrt{(\eta_0^n-\alpha^n) (\eta_0^n-\beta^n)}} . 
\label{zeta-eta}
\end{equation}
The real positive parameters, $\alpha$, $\beta$, and $\gamma$ (where $\alpha>\gamma>\beta>0$), are determined by solving 
\begin{eqnarray}
	R &=& \int_{0}^{\beta}du \frac{\gamma^n-u^n}{%
		\sqrt{(\alpha^n-u^n)(\beta^n-u^n)}} , 
\label{R_abg}\\
	w &=& \int_{\beta}^{\gamma}du \frac{\gamma^n-u^n}{%
		\sqrt{(\alpha^n-u^n)(u^n-\beta^n)}} , 
\label{w1_abg}\\
	w &=& \int_{\gamma}^{\alpha}du \frac{u^n-\gamma^n}{%
		\sqrt{(\alpha^n-u^n)(u^n-\beta^n)}} . 
\label{w2_abg}
\end{eqnarray}
The mapping region in the $\zeta$ plane and that in the $\eta$ plane are shown in Fig.~\ref{Fig_cfm-mapping}. 

Near the edge of a strip at $(x,y)=(R,w)$ (i.e., $\zeta\simeq R+iw$ and $\eta\simeq \gamma$), Eqs.~\eqref{H(z)_Meissner} and \eqref{zeta-eta} are reduced to 
\begin{eqnarray}
	{\cal H}(\zeta) &\simeq& \frac{I_0}{2\pi} \frac{1}{\eta-\gamma} , 
\label{H(z)-eta_edge}\\
	\zeta-(R+iw) &\simeq& -i\frac{n\gamma^{n-1} (\eta-\gamma)^2}{%
		2\sqrt{(\alpha^n-\gamma^n)(\gamma^n-\beta^n)}} , 
\label{z_edge}
\end{eqnarray}
respectively. 
Elimination of $\eta$ in Eqs.~\eqref{H(z)-eta_edge} and \eqref{z_edge} yields 
\begin{equation}
	{\cal H}(\zeta)\simeq \frac{-i\varphi_0}{\sqrt{i(R-\zeta)-w}} , 
\label{H(z)_edge}
\end{equation}
where 
\begin{equation}
	\varphi_0= \frac{I_0}{2\pi} \sqrt{ \frac{n\gamma^{n-1}}{%
		2\sqrt{(\alpha^n-\gamma^n)(\gamma^n-\beta^n)}} } . 
\label{phi-I0}
\end{equation}
Equation~\eqref{H(z)_edge} shows that the sheet current $K_z(y)$ defined as $H_y(R+\epsilon,y)-H_y(R-\epsilon,y)$, where $\epsilon\to +0$, is given by $K_z(y)\simeq 2\varphi_0/\sqrt{w-y}$ and that the perpendicular magnetic field is given by $H_x(R,y)\simeq -\varphi_0/\sqrt{y-w}$. 

Next, we consider the hysteretic ac loss $Q_n$ of polygonally arranged strips carrying an ac transport current $I_0\cos\omega t$ with small amplitude $I_0\ll I_c=2wdj_c$, where $j_c$ is the constant critical current density.~\cite{Bean62} 
As shown in Ref.~\onlinecite{Mawatari06b}, when $I_0\ll I_c$, the ac losses of superconducting strips are directly obtained from the behavior of the magnetic field near the edges of the strip in the ideal Meissner state, as in Eqs.~\eqref{H(z)_edge} and \eqref{phi-I0}. 
The ac loss at the edge $(x,y)=(R,w)$ is given by~\cite{Mawatari06b} $q_0= (4\pi^3/3)\mu_0|\varphi_0|^4/(j_cd)^2$, where $\varphi_0$ is given by Eq.~\eqref{phi-I0}. 
The total ac loss per unit length of polygonally arranged strips is given by $Q_n= 2nq_0$, where the factor $2n$ is the number of edges of the strips. 
The normalized loss $Q_n(I_0)/Q_1(I_0)$ for $I_0\ll I_c$ is given by 
\begin{equation}
	\frac{Q_n(I_0)}{Q_1(I_0)} = \frac{n^3 w^2\gamma^{2n-2}}{%
		(\alpha^n-\gamma^n)(\gamma^n-\beta^n)} , 
\label{Qn/Q1_I0<<Ic}
\end{equation}
where $Q_1$ is the ac loss of a single strip~\cite{Norris70} for $I_0\ll I_c$ and is expressed as 
\begin{equation}
	Q_1(I_0)= \frac{\mu_0}{6\pi} \frac{I_0^4}{I_c^2} 
\label{Q1_I0<<Ic}
\end{equation}
The right-hand side of Eq.~\eqref{Qn/Q1_I0<<Ic} is independent of $I_0$, and is generally expressed as a function of $n$ and $w/R$. 
For $n=1$, Eqs.~\eqref{R_abg}--\eqref{w2_abg} yield $\alpha=\sqrt{w^2+R^2}+w$, $\beta=\sqrt{w^2+R^2}-w$, and $\gamma=\sqrt{w^2+R^2}$, and the right-hand side of Eq.~\eqref{Qn/Q1_I0<<Ic} is reduced to 1, as expected. 
For $nw/R\ll 1$, then $\alpha\simeq R+w$, $\beta\simeq R-w$, $\gamma\simeq R$, and $Q_n/Q_1\simeq n$.

\begin{figure*}[bt]
\includegraphics{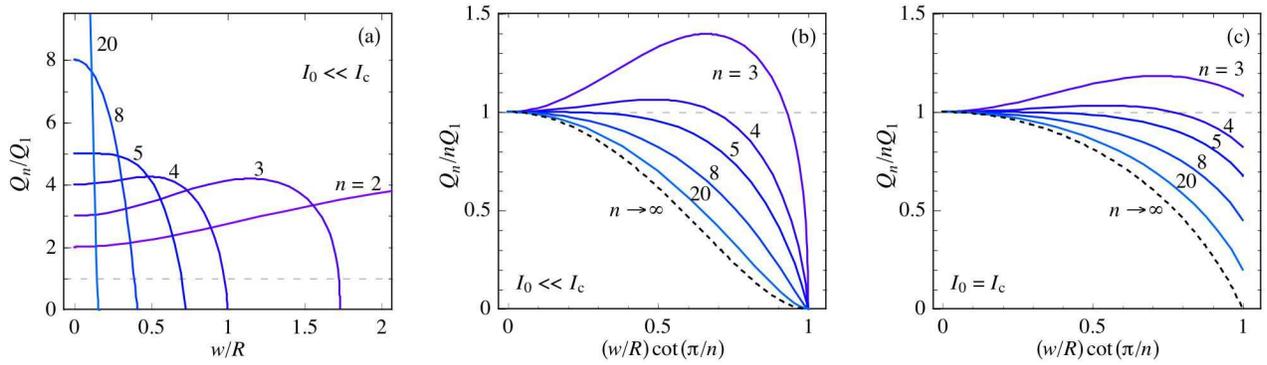}
\caption{(Color online) %
Normalized ac loss $Q_n/Q_1$ as a function of width-diameter ratio $w/R$ for $n=2,3,4,5,8,20$ (solid lines), and $n\to\infty$ (dotted line): 
(a) $Q_n/Q_1$ vs $w/R$ for $I_0\ll I_c$, 
(b) $Q_n/nQ_1$ vs $(w/R)\cot(\pi/n)$ for $I_0\ll I_c$, and 
(c) $Q_n/nQ_1$ vs $(w/R)\cot(\pi/n)$ for $I_0=I_c$. 
}
\label{Fig_Qn-w/R}
\end{figure*}

Figures~\ref{Fig_Qn-w/R}(a) and \ref{Fig_Qn-w/R}(b) show plots of the right-hand side of Eq.~\eqref{Qn/Q1_I0<<Ic} as a function of $w/R$. 
As mentioned in the preceding paragraph, $Q_n/Q_1\simeq n$ for $nw/R\ll 1$, because interaction among multiple strips can be neglected when $nw/R\ll 1$. 
Although $n=2$ strips corresponds to two stacked strips, $Q_2$ monotonically increases with increasing $w/R$. 
For $n=3$ or 4, $Q_n$ nonmonotonically depends on $w/R$, whereas for $n\geq 5$, $Q_n$ decreases with increasing $w/R$ and is less than $nQ_1$. 
When $w/R\to\tan(\pi/n)$, $Q_n$ vanishes, because the edges of neighboring strips approach each other (i.e., the gap between strips approaches zero) and thus the magnetic field component perpendicular to the strips cancels out. 
Note that such behavior of $Q_n\to 0$ for $w/R\to\tan(\pi/n)$ is valid only for the thin-strip limit (i.e., $d/2w\to 0$) and for the small-current limit (i.e., $I_0/I_c\to 0$). 
If finite $d$ or finite $I_0$ is taken into account, then $Q_n$ is finite even when the edges of the strips approach each other.  (The case where $I_0=I_c$ is discussed later.) 

As shown in Fig.~\ref{Fig_Qn-w/R}(b), the dependence of the ac loss of polygonally arranged strips on the strip configuration is clearly revealed by the plots of $Q_n/nQ_1$ vs $(w/R)\cot(\pi/n)$. 
The curves in Fig.~\ref{Fig_Qn-w/R}(b) converge to the dotted line at $n\to\infty$, and the dotted line is simply expressed by 
\begin{equation}
	\lim_{n\to\infty} \frac{Q_n}{nQ_1} 
	= \left(\frac{nw}{2R}\right)^2 \cot^2\left(\frac{nw}{2R}\right) , 
\label{Qn/nQ1_infinite}
\end{equation}
which is derived from Eq.~\eqref{Qn/Q1_I0<<Ic} with $n\gg 1$ and $w/R\ll 1$. 
Note that Eq.~\eqref{Qn/nQ1_infinite} is similar to the ac loss of a coplanar array of infinite strips,~\cite{Mawatari97,Muller97} as follows. 
The ac loss of each strip $Q_{\rm array}$ in an infinite array of coplanar strips is given by~\cite{Mawatari97} 
\begin{eqnarray}
	Q_{\rm array} &=& \frac{\mu_0}{\pi}I_0^2 
		\int_0^1ds (1-2s)\ln\left[1- 
		\frac{\tan^2(s\,\theta I_0/I_c)}{\tan^2\theta}\right] , 
 \nonumber\\
\label{Q_cpl-array}
\end{eqnarray}
where $\theta=\pi w/L$ and $L$ is the periodicity of the coplanar array. 
Equation~\eqref{Q_cpl-array} is valid for $0\leq I_0\leq I_c$, and $Q_{\rm array}$ for $I_0\ll I_c$ is reduced to 
\begin{equation}
	\frac{Q_{\rm array}}{Q_1}\simeq 
		\left(\frac{\pi w}{L}\right)^2 \cot^2\left(\frac{\pi w}{L}\right) , 
\label{Q_cpl-array_small}
\end{equation}
where $Q_1$ is given by Eq.~\eqref{Q1_I0<<Ic}. 
The comparison between Eqs.~\eqref{Qn/nQ1_infinite} and \eqref{Q_cpl-array_small} reveals that polygonally arranged strips for $n\gg 1$ and $w/R\ll 1$ can be regarded as an infinite array of coplanar strips with periodicity $L=2\pi R/n$. 

Equations~\eqref{Qn/Q1_I0<<Ic}, \eqref{Q1_I0<<Ic}, \eqref{Qn/nQ1_infinite}, and \eqref{Q_cpl-array_small} are valid for the small current limit $I_0\ll I_c$. 
The hysteretic ac loss $Q_n$ of polygonally arranged strips at $I_0=I_c$, on the other hand, can be calculated from the geometric-mean distance of the strips~\cite{Norris70} as follows: 
\begin{eqnarray}
	\lefteqn{ \frac{Q_n(I_c)}{\mu_0I_c^2} = \frac{n}{2\pi w^2} } 
\nonumber\\
	&& \times \int_{-w}^{+w}du \int_{-w}^{+w}dv 
		\ln \left[\left| \frac{(R+iu)^n-(R+iv)^n}{R^n-(R+iv)^n} 
		\right|\right] . \hspace{4ex}
\label{Qn_I0=Ic}
\end{eqnarray}
Figure~\ref{Fig_Qn-w/R}(c) shows $Q_n(I_c)/nQ_1(I_c)$ vs $(w/R)\cot(\pi/n)$, where $Q_1(I_c)=\mu_0I_c^2(2\ln2-1)/\pi$ is the ac loss of a single isolated strip~\cite{Norris70} at $I_0=I_c$. 
The fundamental difference between Fig.~\ref{Fig_Qn-w/R}(b) for $I_0\ll I_c$ and Fig.~\ref{Fig_Qn-w/R}(c) for $I_0=I_c$ is the behavior of $Q_n$ for $w/R\to\tan(\pi/n)$; 
even when the edges of neighboring strips approach each other, $Q_n$ of polygonally arranged strips at $I_0=I_c$ does not decrease significantly. 
The $Q_n$ for $I_0\lesssim I_c$ can be reduced by using many narrow strips (i.e., $n\gg 1$ and $w/R\ll 1$) and by arranging the strips so that the edges of neighboring strips approach each other [i.e., $w/R\to\tan(\pi/n)$].

In summary, we theoretically investigated the hysteretic ac loss $Q_n$ of polygonally arranged superconducting strips. 
The $Q_n$ is a function of current amplitude $I_0$, number of strips $n$, and the ratio of the strip width to the diameter of the cylindrical former $w/R$. 
For small current limit of $I_0\ll I_c$, the $Q_n$ is given by Eq.~\eqref{Qn/Q1_I0<<Ic}, and the normalized loss $Q_n/Q_1$ is determined by the strip configuration (i.e., $n$ and $w/R$). 
For $I_0=I_c$, the $Q_n$ is given by Eq.~\eqref{Qn_I0=Ic}. 
Plots of $Q_n/nQ_1$ vs $(w/R)\cot(\pi/n)$, as in Figs.~\ref{Fig_Qn-w/R}(b) and \ref{Fig_Qn-w/R}(c), reveal the effects of the strip configuration on $Q_n$. 
The $Q_n$ of polygonally arranged strips for $n\gg 1$, $w/R\ll 1$, and $I_0\ll I_c$ is given by Eq.~\eqref{Qn/nQ1_infinite}, and is similar to that of an infinite array of coplanar strips that have periodicity $L=2\pi R/n$.

The authors thank M.\ Furuse and H.\ Yamasaki for stimulating discussions.



\end{document}